\documentstyle[prd,aps,epsf]{revtex}
\tighten
\begin{document}
\draft

\preprint{IMPERIAL/TP/95-96/64,HD-THEP-96-49,CGPG-96/12-2}
\twocolumn[\hsize\textwidth\columnwidth\hsize\csname
@twocolumnfalse\endcsname

\title{Non-intercommuting  Cosmic Strings}

\author{Lu\'{\i}s M. A. Bettencourt$^{1,2}$, Pablo Laguna$^3$
and Richard A. Matzner$^{4,5}$}
\address{$^1$The Blackett Laboratory, Imperial College,
London SW7 2BZ, U.K.}
\address{$^2$Institut f\"ur Theoretische Physik, Universit\"at Heidelberg,
Philosophenweg 16, 69120,Germany}
\address{$^3$Department of Astronomy \& Astrophysics and
Center for Gravitational Physics \& Geometry,\\
Penn State University, University Park, PA 16802, USA}
\address{$^4$Center for Relativity, The University of Texas, Austin,
TX 78712, USA}
\address{$^5$Orson Anderson Scholar, Los Alamos National Laboratory, USA
1996-97}
\date{December 9, 1996}
\maketitle

\begin{abstract}

We perform the  numerical field evolution for the collision of two
Abelian type I cosmic strings. We present evidence that,
for collisions at small but characteristic relative
velocities and angles, these cosmic strings {\em do not}
exchange ends and separate. Rather, local higher winding number bound
states are formed close to the collision point, which promote multiple
local scatterings at right angles and prevent intercommutation from happening.
This constitutes the simplest example of the breakdown of the
intercommutation rule, usually assumed in the construction of
effective models for cosmic string network evolution.

\end{abstract}

\pacs{PACS Numbers : 98.80.Cq \hfill 
IMPERIAL/TP/95-96/64, HD-THEP-96-49, CGPG-96/12-2}

\vskip2pc]

Scenarios based on cosmic strings,
formed at a Grand Unified Theory (GUT) phase transition \cite{Kib}
are important candidates to explain the origin of the primordial
perturbations responsible for the formation of structure in the
Universe \cite{Struc}. Cosmic Strings may also be associated with many
other important cosmological phenomena \cite{book}.
After being  formed at a GUT symmetry breaking phase transition,
a network of cosmic strings is thought to
evolve so as to approach a universal scaling behavior, characterized
by a given mean length of string per Hubble volume.
In all implementations to date, this complicated
evolution is assumed to be well described by a
Nambu-Goto action governing the dynamics of each string, together
with a rule for the outcome of the collisions between
them, deduced from the original field theory of which strings are
classical solutions.
Our present ignorance of the details  of GUTs and their  string solutions
makes the latter task impossible. So far
detailed studies of string collisions have been limited to the simplest
field theory exhibiting strings, the Abelian Higgs model.
Cosmic strings formed at a GUT transition may not be
Abelian, even though these solutions are the simplest.

The study of string collisions amounts to solving
an infinite degree of freedom non-linear dynamical system,
which can only be done
numerically. Numerical scattering experiments  in the
Abelian Higgs model, for type II and  global strings,
\cite{She,Mat,Shell},  have  confirmed the usual assumption that  strings
intercommute, i.e., they exchange ends  at every collision.
In this region of parameter space, when the Higgs mass is larger
than the mass of the gauge field, the interactions between
two strings with the same orientation are  repulsive \cite{Jac,BR},
leading to their separation after the collision.

Type I strings are more interesting  because the static potential
between them is always attractive \cite{Jac,BR}. As a consequence,
higher winding number bound states can be formed.
In particular these bound states
prevent an ordered Abrikosov lattice from existing in laboratory
experiments involving type I superconductors. Nevertheless,
a network of type I strings  is thought  to be viable in the early Universe
as long as the string density at formation is sufficiently low \cite{BR}.
All numerical studies  concerning the outcome of
type I string collisions performed to date \cite{LM} were  targeted at showing
that, at high approach center-of-mass velocities ($v=0.75$ with $c=1$),
two high winding number strings will form a bridge of lower winding
number connecting them. This bridge then  grows, promoting  the peeling of
the original high winding number configurations onto lower ones \cite{LM}.
\begin{figure}
\centerline{\epsfxsize=3.2in\epsfbox{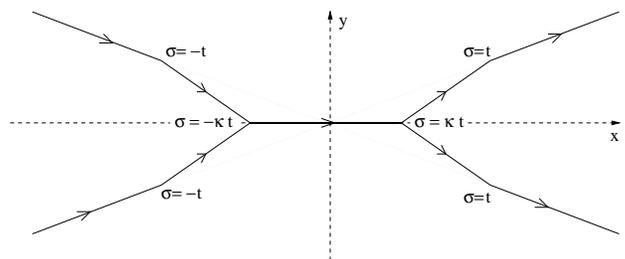}}
\caption{The {\it zipper} configuration. The higher
winding number bridge (bold) grows along the strings if the collision
angle and velocity are  small. $\sigma$ is
the co-ordinate along the string and $\kappa$ is the {\it zipper} velocity.}
\label{fig1}
\end{figure}

At a phase transition, one expects to form predominantly unit
winding number strings.
Consequently, it is interesting  to investigate the converse process, i.e.,
if and when the collision of two strings can result in the formation of
a higher winding number bridge between them
and for what range of model and dynamical parameters.
In this Letter, we present evidence for the existence of such bound states.
They constitute the simplest example of the breakdown of the intercommutation
rule. We expect an initially growing bridge solution, or
{\it zipper} (see Fig.~\ref{fig1}), to exist for small
approach velocities $v$ and angles $\alpha$.\cite{BK}
The meaning of small depends, in turn,  on the ratio of the Higgs to
gauge masses in the model.
We define the  Abelian Higgs model by the Lagrangian
\begin{eqnarray}
{\cal L} = -{1 \over 4 } F_{\mu \nu} F^{\mu \nu} + {1 \over 2} \vert
(\partial_\mu + i e A_\mu) \phi \vert ^2 -{\lambda \over 8} (\vert \phi
\vert ^2 - \eta^2)^2.
\label{e1}
\end{eqnarray}
Type I strings  correspond to $\beta \equiv e/\sqrt{\lambda} > 1$,
and type II to $\beta < 1$.
For $\beta=1$, the static attractive and repulsive potentials
exactly cancel, resulting in no interactions between
vortices in 2D \cite{Jac,BK}.

Under the simple assumption that the {\it zipper} grows
with {\it constant} velocity, $\kappa$,
it is possible to find a solution for the Nambu-Goto equations
relating  $\kappa$ to the collision parameters \cite{BK},
$\kappa = (\xi -1/\epsilon)/(\epsilon -\xi)$,
where $\epsilon > 1$ is the ratio of twice the energy per unit length
of a unit winding number, $n=1$, string to that of an $n=2$ string, and
$\xi = \cos(\alpha/2) \sqrt{1- v^2}$.
A growing  {\it zipper} can only exist for $\xi > 1 / \epsilon$,
i.e., for small enough angles $\alpha$ and/or small approach velocities $v$.
If the {\it zipper}'s growth $\kappa$ is constant, it could in
principle continue forever. However, this is a  simplistic scenario.
It not only assumes that the strings away from the {\it zipper}
remain straight and travel at constant velocity, but
this picture also neglects the attractive
interaction between string segments close to junctions.
In realistic circumstances, a more complicated motion should take place,
namely changes in the relative velocity and/or angles between the
interacting strings.
Our results  show  that as long as the angle at the junction is
small enough,  optimal conditions exist for the {\it zipper} to grow.

In our numerical experiment,
we chose as initial conditions two
$n=1$ straight strings at a relative angle $\alpha$ and
approaching each other with initial center of mass velocity $v$.
The dynamical parameter space is given by these two
variables and the model parameter space by
$\beta$. We keep $\beta=2$ throughout.
The boundary conditions correspond to  boosted, unperturbed free vortex
configurations.  A typical set of events is presented in Fig.~2.
Initially, we observe the two strings approaching each other, Fig.~2a,
and  the colliding segments at the center re-emerging at $90^\circ$, Fig.~2b.
This is expected because, at very small separations,
the interactions between  colliding string segments are
unimportant \cite{Jac}  and the
geometry of the collision can  be understood on topological
grounds \cite{Manton}.  Other segments of string,  that have not collided
head-on but have impact parameter smaller than twice its  width, are
also affected by the interactions. Their orbits are curved inwards,
in a manner similar to what has been observed in 2D studies \cite{Rebbi}.
The result of the first scatter is the configuration of Fig.~2c.
The two strings appear to have been twisted relative to each other.
The  velocity of the string segments in the central region
is now predominantly orthogonal to the original one.
At the collision point, in particular, the velocity
has no component in the original incoming direction.
Globally, the magnitude of the velocity  in the orthogonal direction
diminishes as one goes outwards along the strings, while the component
in the original direction increases  approaching
its asymptotic value on the boundaries.
Meanwhile, the segments with original impact parameter smaller than
twice the width of the string remain under the effect of the interactions.
The attractive forces  are
strongest at the center and bring together the string segments once
more, now more softly. These then scatter again at $90^\circ$ but lack
the initial large kinetic energy to re-emerge as individual $n=1$
segments, Fig.~2d. While these events take place in the central region,
the  segments of string immediately further out orbit slowly around each other
and fall  inwards. This  configuration creates optimal conditions for
the $n=2$ {\it zipper}  to form.

The {\it zipper} then rapidly grows outwards at a
large $\kappa$ speed, as seen in Figs.~2d and 2e.
Figure~2e shows the
configuration when the kinks at $\sigma = \pm t$
(see Fig. 1) reach the boundaries of the computational domain.
After that point, because of the boundary conditions
(boosted, unperturbed strings), the opening angle of the {\it zipper}
increases.
The angle formed by the two $n=1$ segments of string at the $n=2$ junction
in Fig.~2e is $\sim 60^\circ$.
This is larger than the small angle necessary to sustain the
growth of the {\it zipper}. The motion at the junctions decelerates and
ultimately stops.
The result is an  approximately static
$n=2$ bridge joining the original $n=1$ strings. It is important to
point out that because  the
$n=2$ bridge has reached an approximately static state,
there is {\it no memory} of the origin  of its two $n=1$ components.
At this point, the unperturbed vortices at the edge of the computational
domain act as the dominant force determining the dynamics of the {\it zipper}.
Although in the present case, this is the result of our boundary conditions,
in a realistic string network evolution, the forces acting on the
{\it zipper} are due the dynamics of string segments
much longer than the effective length participating in the collision.
The consequence is  the unpeeling of the $n=2$ bridge,  in such a way that the
two original $n=1$ strings re-emerge from the scatter
as if they had gone through each other {\it without}
having exchanged ends (see  Fig.~2f). There is effectively {\it no}
intercommutation!

We verified that intercommutation always takes place for
type II strings, in identical circumstances \cite{Movies}. 
The onset of the {\it zipper} was also 
tested by performing several string collision
simulations for different computational domain sizes.
The  results presented in Fig.~2 were obtained on a
$128^2\times 256$ mesh. We have performed identical evolutions on meshes
of up to $180^2\times 810$,
maintaining the lattice spacing constant.
The result of increasing the size of the computational domain was {\it only}
a longer growth of the {\it zipper}. This is  because the arrival of the 
string kinks at the boundaries, and the subsequent increase in the opening 
angle, is more delayed the larger the computational domain.
Moreover, we explicitly observed the transition from the sequence of events
involving the {\it zipper} to the usual intercommutation  by increasing
the initial approach velocity and/or  the angle $\alpha$.
For an initial velocity $v=0.1$,
two strings effectively do not intercommute for angles smaller
than $25^\circ$.
For larger approach velocities the angle for which the transition
between the two outcomes happens is smaller, e.g., about
$\alpha = 20^\circ$ for $v=0.15$.
These results are in good agreement with the original estimates for the
{\it zipper} formation \cite{BK}.
The main source on uncertainty in \cite{BK} was that, due to
the extended nature of the strings, it was difficult to
estimate how much kinetic energy participates in the collision.
The amount of string carrying relevant kinetic energy was then parameterized
by a characteristic length, $l_{\rm eff}$.
Fig.~2, makes it  possible to measure the length
of string in the {\it zipper}. In units of the width of the
string, it is about 30. Adopting this  value for $l_{\rm eff}$
we find  transient velocities and angles very similar to the ones
measured here. For $v=0.15$ the resulting angle
indicates that a smaller length of string (about $l_{\rm eff} = 20$)
carries relevant kinetic energy.
This is compatible with a faster collision and the observed extent of the
{\it zipper}.

In conclusion, our numerical experiments indicate that
the outcome of the collision of two type I cosmic strings with
sufficiently low scattering energy at small angles follows two stages.
First, the two strings collide, and, due to the
attractive forces between them, after a few  local scatters at right
angles,  settle down to a local bound state, the {\it zipper}.
Second, the {\it zipper} grows
under the effect of the attractive interaction between segments of string
at its two ends. The final fate of the {\it zipper} depends
on whether the condition for {\it zipper} formation,
$\xi > 1 / \epsilon$, is violated (large collision angle and/or velocity).
In our simulations, 
this violation was the result of the boundary conditions,
but in a string network evolution it could be due to the string
dynamics itself.
In any case, if a {\it zipper} stops growing,
its free ends will pull the strings
apart and unzip the $n=2$ bridge in  the energetically most favorable
way, namely the configuration that minimizes the overall string length.
In our simulations, such configuration is that in which strings {\it do not}
exchange ends. Whenever the kinetic energy involved
in the collision is larger than the attractive potential between the strings
the {\it zipper} cannot settle in and usual intercommuting takes place.

In closing, we note that the results obtained above correspond to very
mildly type I strings, $\beta= 2$. It is very interesting
that large relative couplings are not necessary for the {\it zipper} to form.
The extent of the
space of $v$ and $\alpha$ for which a {\it zipper} and no intercommutation
will occur is  dependent on the  relative strength of the
interactions $\beta$. The fraction of non-intercommuting collisions is
a measure of how strongly type I a theory is.
Type I theories are thought to display first order transitions,
which in standard cosmological scenarios  seem to be  necessary  at the GUT
scale in order to account for the baryonic asymmetry of the Universe.

The consequences of the breakdown of the intercommutation rule
for cosmic string network dynamics can be very
important. Intercommutation, leading to the formation of small loops from
long strings that can subsequently decay,
is the feature that makes cosmic strings viable cosmologically in contrast
to other  topological defects.
Loop production and decay
provides the string network with an effective  means of dissipation,
preventing it from  dominating the energy density of the Universe.
In practice, given that $v$ and $\alpha$ must be small (the
string coherent velocity measured in  network evolutions, is
$v\sim 0.15$ \cite{VV}), it seems
likely that in a network of type I strings
only a fraction of all collisions will result in no
intercommutation. Then the string network will possess globally a less
effective mechanism of producing loops and thereby losing energy.
A slower approach to the expected universal scaling regime
would probably follow, allowing for more string to be present at
later times. It would be interesting to determine whether such a string network
would permit viable structure formation scenarios, and, in what sense
their predictions would be different.

We thank Brandon Carter and Tom Kibble for their
enquires and suggestions.
LMAB thanks JNICT-{\it Programa Praxis XXI},
contract No.~BD~2243/92~RM  and the DFG for financial support. 
PL was partially supported by
NSF grants PHY~96-01413, 93-57219~(NYI).
RAM was partially supported by NSF grant PHY 93-10083.

\newpage
\onecolumn
\vspace{2cm}
\begin{figure}
\label{st1}
\centerline{\epsfxsize=5.5in\epsfbox{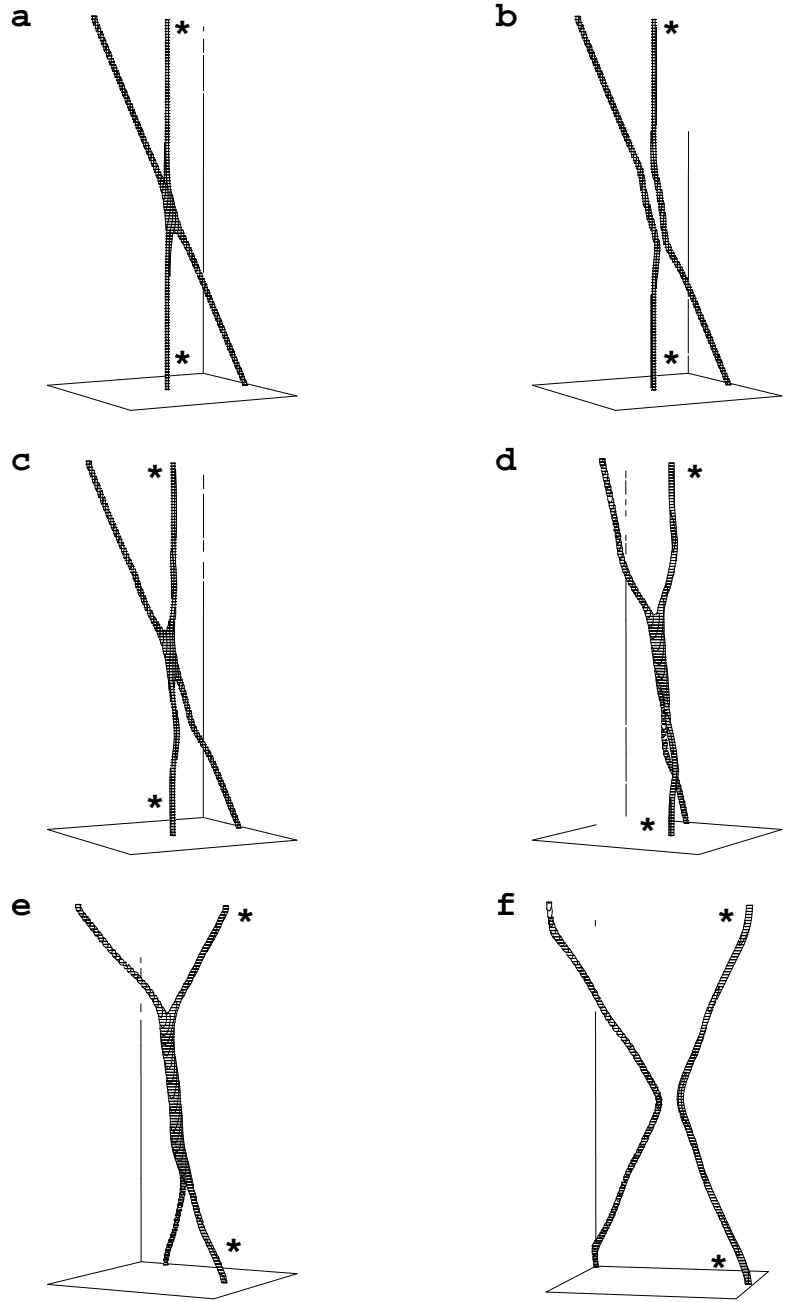}}
\vspace{2cm}
\caption{Snapshots of the numerical evolution of the scattering of
two type I cosmic strings with parameters $v=0.1$, $\alpha=25^\circ$
and $\beta=2$. Contour levels represent $65\%$ of the energy at the
core of the strings. The viewing angles, with respect to the colliding
velocity direction, are $30^o$ for (a-c), $60^o$ for (d) and (e), and
$85^o$ for (f). The stars label the two ends of one of the two original 
strings.}
\end{figure}

\end{document}